\documentclass[fleqn,usenatbib]{mnras}

\usepackage{newtxtext,newtxmath}

\usepackage[T1]{fontenc}
\usepackage{ae,aecompl}

\usepackage{graphicx}	
\usepackage{amsmath}	
\usepackage{amssymb}	
\usepackage{multicol}
\usepackage{enumitem}

\title{Connecting HL Tau to the Observed Exoplanet Sample}
\author[C. Simbulan et al.]
{Christopher Simbulan,$^{1}$
Daniel Tamayo,$^{2,3,4}$
Cristobal Petrovich,$^{3,5}$
\newauthor
Hanno Rein,$^{2,1}$
Norman Murray$^{3}$\\
$^1$ Department of Astronomy and Astrophysics, University of Toronto, Toronto, Ontario, M5S 3H4, Canada\\
$^2$ Department of Physical and Environmental Sciences, University of Toronto at Scarborough, Toronto, Ontario M1C 1A4, Canada\\
$^3$ Canadian Institute for Theoretical Astrophysics, 60 St. George St, University of Toronto, Toronto, Ontario M5S 3H8, Canada\\
$^4$ Centre for Planetary Sciences Fellow\\
$^5$ Canadian Institute for Theoretical Astrophysics Fellow
}

\date{Accepted XXX. Received YYY; in original form ZZZ}

\pubyear{2015}

\begin{document}
\label{firstpage}
\pagerange{\pageref{firstpage}--\pageref{lastpage}}
\maketitle

\begin{abstract}
The Atacama Large Millimeter/submilimeter Array (ALMA) recently revealed a set of nearly concentric gaps in the protoplanetary disk surrounding the young star HL Tau. If these are carved by forming gas giants, this provides the first set of orbital initial conditions for planets as they emerge from their birth disks. Using N-body integrations, we have followed the evolution of the system for 5 Gyr to explore the possible outcomes. We find that HL Tau initial conditions scaled down to the size of typically observed exoplanet orbits naturally produce several populations in the observed exoplanet sample. First, for a plausible range of planetary masses, we can match the observed eccentricity distribution of dynamically excited radial velocity giant planets with eccentricities $>$ 0.2. Second, we roughly obtain the observed rate of hot Jupiters around FGK stars. Finally, we obtain a large efficiency of planetary ejections of $\approx 2$ per HL Tau-like system, but the small fraction of stars observed to host giant planets makes it hard to match the rate of free-floating planets inferred from microlensing observations. In view of upcoming GAIA results, we also provide predictions for the expected mutual inclination distribution, which is significantly broader than the absolute inclination distributions typically considered by previous studies. 
\end{abstract}




\section{Introduction}
\par A wealth of discoveries over the last two decades has generated an extensive and diverse exoplanetary database \citep[see, e.g.,][and references therein]{Udry07, Howard13, Fabrycky14}.  Perhaps most immediately striking is the fact that many systems are much more dynamically excited than the Solar System \citep{Winn15}.  Additionally, observations reveal that $\sim$ 1\% of F, G, and K dwarf stars host hot Jupiters at short orbital periods \citep{mayor2011harps, Howard12, wright2012}, where it would seem difficult for them to form in situ\citep{Bodenheimer00}.  Microlensing surveys have further inferred a frequency of $\sim$ 1.3 free-floating Jupiter mass planets per main sequence star \cite{sumi2011}, which they interpret as planets ejected from their birth system.  The above constraints all point to tumultuous dynamical histories for these planetary systems.

\par Several mechanisms have been proposed to dynamically excite systems, leading to planetary scatterings, ejections, and at high enough eccentricity, circularization through tides to form hot Jupiters. First, under the right conditions, direct interactions with the protoplanetary disk can grow planetary eccentricities \citep{artymowicz1992, goldreich2003, papaloizou2001, ogilvie2003}.  Another possibility is for mean-motion resonances to raise eccentricities under the action of disk or planetesimal-driven migration \citep[e.g.,][]{lee2002,ReinPapaloizouKley2009}, or for secular interactions with stellar \citep{wu2003,fabrycky2007} or planetary \citep{Naoz11, Wu11, Petrovich15} companions to drive interior planets onto highly elliptical paths. Alternatively, planets could form in nearly circular but extremely compact configurations that over time will lead to dynamical excitation through gravitational scattering \citep{rasio1996, weidenschilling1996, lin1997}.

\par Previous studies involving two planets of equal mass \citep{ford2001dynamical,ford2003,ford2008} performed simulations showing that planet-planet scattering can dynamically excite systems, but not enough to match the observed distribution. However, systems containing two planets with unequal masses, or systems containing three or more planets, have been able to produce distributions closer to the observed sample \citep{marzari2002,ford2003,ford2008,Chatterjee08,Juric08}.

\par One important challenge for these studies is that the various dynamical phenomena above depend on particular disk conditions and on the initial orbital configurations following planet formation.  By varying parameters to match today's observed exoplanet distribution \cite{Chatterjee08} and \cite{Juric08} infer an appropriate range of initial conditions, but this inverse problem inevitably leads to degeneracies among model parameters. The ideal scenario would be to instead determine initial conditions of planet formation directly through observation and evolve them for Gyr timescales to compare the results to the comparably aged exoplanet sample.

\par Unfortunately observations of forming planetary systems are difficult due to their small angular scale, the stage's short (Myr) duration, and the high levels of dust extinction during this phase.  However, the Atacama Large Millimeter/submilimeter Array (ALMA) is revolutionizing the field through its ability to probe optically thin millimeter-wavelength continuum emission at unprecedented AU-scale resolution.

\par In October 2014, \cite{Brogan15} obtained an image of the young star HL Tauri (hereafter HL Tau), revealing a set of nearly concentric gaps (Fig.\:\ref{gaps}) in the dust distribution. More recently, \cite{Yen16} have also observed gaps in the gas at coincident locations. Several mechanisms have been proposed to explain the origin of the gaps, such as clumping \citep{lyra2013}, magnetic instabilities \citep{pinilla2012}, and rapid pebble growth due to condensation fronts \citep{zhang2015}. However, arguably the most intriguing explanation is that the gaps are carved by forming giant planets \citep{Brogan15,Dipierro15}

Theoretical studies have long shown that giant planets are expected to carve out gaps in the surrounding gas disk during their formation \citep[e.g.][]{lin1986}. One complication is that the continuum emission in the ALMA image probes not the gas, but rather $\sim$millimeter-sized dust grains in the disk, which are additionally subject to aerodynamic forces. Two-phase (gas+dust) hydrodynamical codes have been developed to trace dust behavior, and find that planets as small as $\sim 1$ Neptune mass are capable of opening a gap in the surrounding dust \citep{Fouchet10}.

But if Neptune-mass or larger planets orbit in the observed gaps, this has important implications for the stability of the system. \cite{Tamayo15} showed through N-body integrations that if planets orbit in each of the observed gaps, the current masses must be $\lesssim 1$ Saturn mass in order to ensure orbital stability over the lifetime of the system. However, even these stable cases are only transient.  As the planets continue to accrete gas from the surrounding disk, and on galactically significant timescales of Gyrs, the systems will destabilize, undergoing close encounters that eject planets to interstellar space while leaving the remaining planets on inclined, eccentric orbits \citep{rasio1996,ford2008,Chatterjee08,Juric08}

\par Given that the HL Tau system potentially provides us the first set of initial conditions as giant planets emerge from their birth disk, we undertake an analysis of the potential scattering outcomes.  We then compare the results following several Gyr of dynamical evolution to the observed exoplanet sample.

We begin by describing our numerical methods and sets of initial conditions in Sec.\:\ref{methods}. We then compare our resulting eccentricity distribution across sets of initial conditions (Sec.\:\ref{sensitivity}) and to the observed radial velocity sample (Sec.\:\ref{comptoobserved}). In Sec.\:\ref{massdistribution} we explore the effect of adding a mass distribution to the planets. In Sec.\:\ref{inclinations} we make predictions for the expected distribution of mutual inclinations between planets, and in Sec.\:\ref{freefloating} investigate the production of free-floating planets. Finally, in Sec.\:\ref{hotjupiters}, we explore the production rate of Hot Jupiters, and their expected obliquity distribution.

\section{Methods} \label{methods}

\subsection{The HL Tau System}
\label{hltausystem}

\begin{figure} 
\includegraphics[width=\columnwidth]{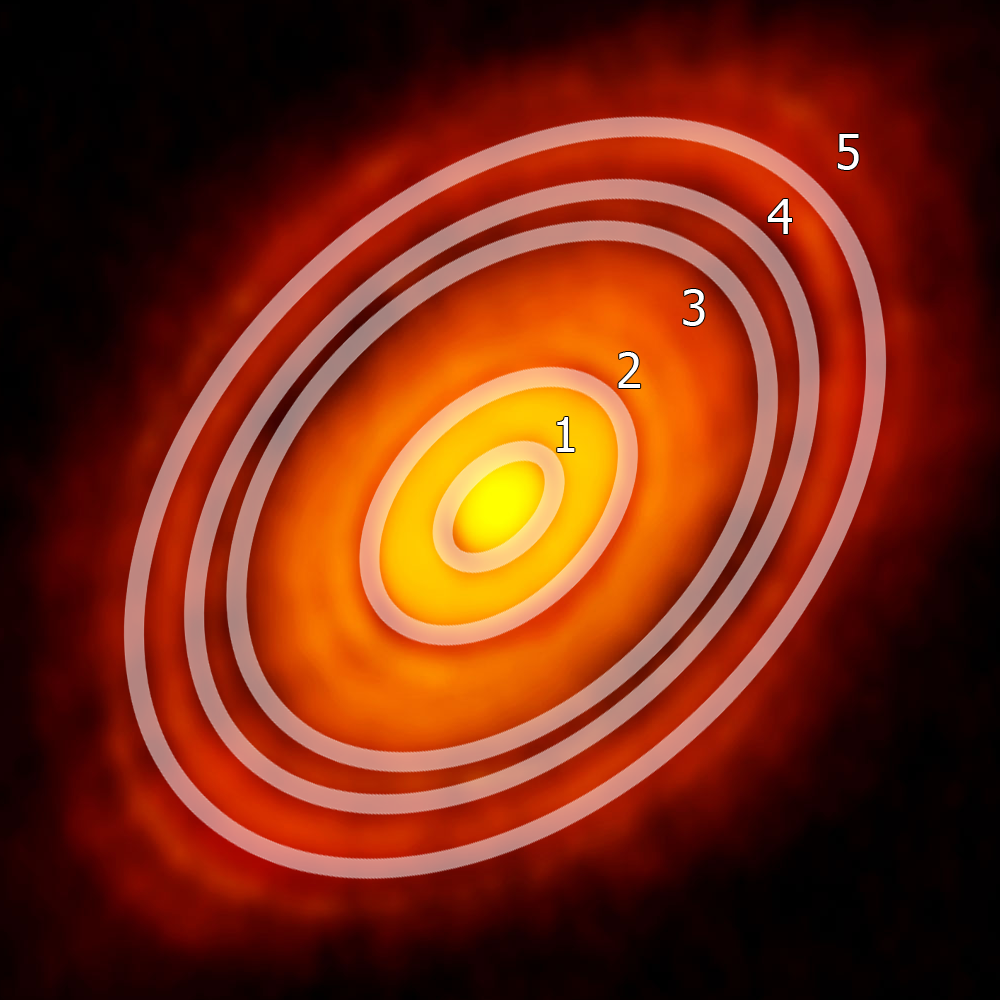}
\caption{ALMA's 233 GHz continuum image of HL Tau (\url{http://www.eso.org/public/news/eso1436}), with white ellipses overlaid on the five most prominent dark bands and labeled 1 to 5.
For scale, gap 5 lies $\approx$ 91 AU from the central star.
}
\label{gaps}
\end{figure}

\cite{Brogan15} identified five major gaps in the HL Tau disk, which we label gaps 1-5 (see Fig.\:\ref{gaps}).  Planets orbiting in the inner two gaps are well enough separated from other gaps that they are dynamically stable on Gyr timescales.  The stability of the system therefore hinges on any planets orbiting in the outer gaps \citep{Tamayo15}.

\cite{Akiyama16} argue that each of the gaps' widths should be comparable to its progenitor planet's Hill sphere, inferring masses consistent with giant planets\footnote{\cite{Akiyama16} do not consider Gap 5, but their same reasoning would apply to this gap given its comparable width to the other ones}. Their inferred masses are likely overestimates, since the gap widths observed in the dust distribution from hydrodynamic simulations including both gas and dust are generally several Hill radii \citep[e.g.,][]{Fouchet10, Dipierro15}; however, many disk parameters remain uncertain and several ingredients are important to the disk physics and radiative transfer \citep[e.g.,][]{Pinte16, Jin16, Li16}.  

We first considered a case with 5 planets in each of gaps 1-5. The initial conditions for this experimental setup were taken from the gap locations reported by \cite{Brogan15} and can be found in the top row of Table \ref{tableconditions}.  A more detailed description of the numerical setup for all suites of simulations can be found in Sec.\:\ref{setup}.

Alternatively, one might interpret the two closest gaps (3 and 4) as due to a single planet, with the intervening material representing dust that is co-orbital with the planet.  This scenario was modeled with hydrodynamic simulations including gas and dust by \cite{Dong15} and \cite{Dipierro15}, inferring a mass for this planet of $\sim 0.5 M_J$.  We therefore ran an additional suite of simulations with planets in each of gaps 1, 2 and 5, and a single planet at the location of the brightness peak between gaps 3 and 4, reported by \cite{Brogan15} (see the last two rows in Table \ref{tableconditions}).

Several authors \citep[e.g.,][]{Brogan15} have noted that the orbital periods at the locations of the various gaps in the HL Tau disk form near-integer (i.e., resonant) period ratios. The ALMA resolution, together with our ignorance of where within the gap the planet orbits makes it impossible to directly determine whether this is the case.  However, \cite{Tamayo15} suggested that {\it resonantly} interacting planets experiencing damping from the surrounding disk should develop eccentricities of a few percent, which was consistent with the gap offsets later reported by \cite{Brogan15}.  Nevertheless, for generality, we chose to simulate (for each of the four and five planet scenarios) both a case where planets are initialized with random angles, and one where the planets were initially damped to the center of the nearest resonance (see Sec.\:\ref{sec:initialconditions} for details).

Previous works \citep[e.g.,][]{Juric08, Chatterjee08} have found that suites of unstable planetary systems with different initial conditions tend to converge to the same final orbital distribution, as the chaotic interactions tend to erase a system's memory of its initial conditions.  One might therefore hope that our four cases summarized in Table \ref{tableconditions} would converge to the same distribution, obviating the need for precise initial conditions.  We examine this hypothesis in Sec.\:\ref{results}.

One caveat to the above-mentioned convergence of outcomes is that, depending on the number of planets and their masses, the HL Tau system may be {\it stable} over Gyr timescales.   In particular, if we ignored the largely dynamically decoupled inner two planets, our four-planet case reduces to an effective two-planet system.  In this case we can apply the analytic Hill stability criterion \citep{Marchal82, Gladman93} to find that if the sum of the two planets' masses are $\lesssim 2 M_J$, the planets cannot undergo close encounters.  Adding a nearby third planet (i.e., the five-planet case) breaks this restriction, and leads to eventual instability.  In all suites of simulations, we adopted planetary masses that would render the systems unstable, translating to higher masses in the four-planet case (see Sec.\:\ref{setup}).

\begin{center}
\begin{table*}
\begin{tabular}{|l|c|c|c|c|c|c|c|c|c|}
\hline \hline
Case                  & $a_1$   & $a_2$   & $a_3$   & $a_4$   & $a_5$   & $e_{\rm inner}$ & $e_{\rm outer}$   & i                                  & f                                    \\ \hline \hline
5 Planet Nominal & 13.2 AU & 32.3 AU & 64.2 AU & 73.7 AU & 91.0 AU & 0             & 0             & $U{[}0^{\circ} , 1^{\circ}{]}$ & $U{[}0^{\circ} , 360^{\circ}{]}$ \\ \hline
5 Planet Resonant     & 13.2 AU & 32.3 AU & 64.4 AU & 74.8 AU & 90.8 AU & $\sim10^{-5}$ & $\sim10^{-3}$ & $\sim10^{-2}$                      & Resonant                             \\ \hline

4 Planet Nominal & 13.2 AU & 32.3 AU & 69.0 AU & 91.0 AU & N/A     & 0             & 0             & $U{[}0^{\circ} , 1^{\circ}{]}$ & $U{[}0^{\circ} , 360^{\circ}{]}$ \\ \hline

4 Planet Resonant     & 13.2 AU & 32.3 AU & 69.1 AU & 90.8 AU & N/A     & $\sim10^{-5}$     & $\sim10^{-3}$     & $\sim10^{-2}$                      & Resonant                             \\ \hline
\end{tabular}
\caption{Initial orbital elements for the different cases considered (semimajor axes $a$, eccentricities $e$---separated between the dynamically largely decoupled inner two planets, and the more strongly interacting outer planets---inclinations $i$ and phase along the orbit $f$). Between the resonant and nominal cases, only the outer planets' semi-major axes differed. See the main text for a discussion of how the resonant cases were initialized. $U[]$ in the columns for inclination and phase represents drawing from a uniform distribution in the quoted range.}
\label{tableconditions}
\end{table*}
\end{center}

\subsection{Numerical Integrations} \label{setup}

Our simulations made use of the open-source N-Body package REBOUND \citep{rein2012}.  In particular, all integrations were performed with the adaptive, high-order integrator IAS15 \citep{rein2015ias15}. Any additional forces were incorporated using the REBOUNDx library\footnote{\url{https://github.com/dtamayo/reboundx}}.

\par One hundred integrations were performed for each case listed in Table \ref{tableconditions}, and each system was integrated for 5 Gyr. To limit the parameter space, we assigned all planets the same mass and radius.  In all runs, the radius was set to a constant value of $\approx 1$ Jupiter radius ($R_J \equiv 71492$km). We note that our results in Sec.\:\ref{results} are insensitive to the exact radius adopted, since collisions are rare\footnote{At the large orbital distances in the HL Tau system, where the orbital velocities are much smaller than the escape velocities from the planets' surfaces, scattering events dominate collisions\citep[e.g.,][]{Petrovich14}.}.

\par All planets were begun as proto-giant planets with $M_0 = 10 M_{\oplus}$. At approximately this value, one expects the mass accretion to transition to exponential growth \citep[e.g., ][]{Pollack96}. At some point this accretion must halt, though the exact mechanism by which this occurs is unclear \citep[e.g., ][]{Morbidelli14}. We therefore made the simple choice to grow the mass with the {\tt modify\_mass} implementation in REBOUNDx, using a time-dependent e-folding timescale $\tau_M$ that keeps the masses in the planetary regime,

\begin{equation} \label{taum}
\tau_M = \tau_{M0}e^{(t/\tau_{\rm Disk})},
\end{equation}
where $\tau_{M0}$ is a constant, and the exponential factor is meant to qualitatively capture the dispersal of the protoplanetary disk on a timescale $\tau_{\rm Disk}$. We set $\tau_{\rm Disk}$ = 3 Myr, which corresponds the median timescale observed for the disappearance of protoplanetary disks \citep{Haisch01}. Taking expected accretion timescales $\tau_{M0}$ \citep[e.g.][]{Ida04} typically yield planetary masses that are too high \citep[e.g.,][]{Szulagyi14}.  In our five-planet cases, we therefore simply tuned $\tau_{M0}$ to 0.85 Myr, which translated to final planetary masses of $\approx 1.1$ Jupiter masses ($M_J$).  This provided a typical observed mass for giant planets, and was large enough for all systems to go unstable. 

We note that once planets reach a mass at which pairs of adjacent planets are Hill unstable, the instability happens swiftly on a few conjunction timescales \citep[e.g.,][]{Gladman93}. This means that the particular mass-growth prescription adopted does not have a large effect on the results. We verified this by running an additional suite of simulations mimicking the 5-planet nominal initial conditions, but with no mass-growth prescription---planets were simply initialized with their final mass of $1.1$ Jupiter masses. The resulting cumulative eccentricity distribution was statistically consistent with having been drawn from the same distribution as the simulations with the mass-growth prescription above.

In the four-planet case, however, $1.1 M_J$ was not large enough to render the majority of systems unstable. Since we later aim to compare to both the five-planet case and the observed exoplanet sample, we therefore decreased $\tau_{M0}$ to 0.60 Myr, corresponding to final masses of $4.7M_J$.  This rendered most systems unstable. This promotes a loss of memory of initial conditions through chaotic interactions \citep[e.g.,][]{Juric08}; in this way, a single set of initial conditions can hope to reproduce the wider exoplanet sample.

To simulate dissipation from the disk, we followed \cite{Tamayo15} and used the {\tt modify\_orbits\_forces} implementation in REBOUNDx to apply eccentricity damping at constant angular momentum \citep{Papa00}.  As above, we modify the e-folding timescale $\tau_e$ to qualitatively capture the disk dispersal,

\begin{equation}
\tau_e = \tau_{e0}e^{(t/\tau_{\rm Disk})},
\end{equation}
where we set $\tau_{e_0}$ = 1 Myr ($\sim$ $10^3$ orbits) for all simulations. The above effects were turned off after 40 Myr, at which point their effect is negligible.

In a more complete physical picture, the details of the eccentricity damping would change as the planets grow, clear a gap in the disk etc. \citep[see, e.g., the review by][]{Baruteau14}. While eccentricity damping can stabilize planetary systems against slow chaotic diffusion \citep{Tamayo15}, once adjacent planets reach a mass at which they are Hill unstable, the instability is so violent that no reasonable amount of eccentricity damping will stabilize the system. We ran a copy of the 5-planet nominal suite of simulations but with $\tau_{e0}$ a factor of ten larger, and obtained statistically consistent results.

As the simulation progressed, planets were removed if they reached a distance of 1000 AU from the star. All collisions between bodies were treated as perfectly inelastic mergers.

There is considerable uncertainty in the mass of the central star. Estimates range from $\sim 0.55$ $M_\odot$ (solar masses, \citealt{Beckwith90}) to $\sim 1.3 M_\odot$ \citep{Brogan15}. For simplicity, we adopted a stellar mass of $1 M_\odot$. In the absence of additional forces and collisions, the point-particle N-body problem can be non-dimensionalized by expressing all the masses in units of the central mass, and time in units of the innermost orbital period. We argued above that the additional effects have little impact on the outcomes, and as we will see below, collisions are largely negligible. Thus, adopting a different stellar mass closely approximates proportionately adjusting the masses of the planets and the timescales from our reported values.

\par For computational reasons, if planets reached separations $<0.2$ AU from the central star, we saved a checkpoint of the simulation, merged the planet with the star, and continued the integration. We present the results of these simulations in Sec.\:\ref{results}. In a separate investigation, we subsequently loaded the above-mentioned checkpoints where planets reached 0.2 AU, and continued these integrations with additional short-range forces (general relativity corrections, tidal precession) taken into account.  We discuss those results in Sec.\:\ref{hotjupiters}.

Each integration required a different amount of time depending on the closeness of planetary encounters, but each 100-simulation case we executed required $\sim 3 \times 10^4$ hours on older Intel Xeon CPUs (E5310, 1.6 GHz).

\subsection{Initial Conditions} \label{sec:initialconditions}


\par As discussed above, the semimajor axes for the nominal cases were taken from the values reported by \cite{Brogan15}.  The eccentricities were initialized to zero, while the inclinations were drawn from a uniform distribution in the interval $[0,1^\circ]$.  All remaining orbital angles were randomly drawn from the interval $[0,360^\circ]$.

For the resonant cases, we note that capture into resonance depends on the relative rate at which pairs of planets migrate toward one another in the disk \citep[e.g., ][]{Quillen06}, which in turn remains uncertain \citep[e.g., ][]{Baruteau14}. Determining whether capture into resonance is likely in the case of HL Tau is beyond the scope of this paper---we simply place planets in resonance to see whether such configurations lead to qualitatively different outcomes. We will find in Sec.\:\ref{sensitivity} that the different initial configurations lead to consistent outcomes, obviating some of these concerns. In any case, the steps below should be seen as a simple numerical procedure to obtain resonant chains, rather than reflecting physically plausible parameters from disk migration.

We began with planets initialized as in the nominal case, but we moved outward and slightly separated the outer three planets,  We then added semimajor axis damping to the outermost planet following the prescription of \cite{Papa00} that is implemented in the {\tt modify\_orbits\_forces} routine in REBOUNDx.  The outermost planet then migrated inward until it captured into the desired resonance nearby the observed separations, at which point the pair of planets continued migrating inward. In the 5-planet case, this resulted in the outer two planets locking into a 4:3 resonance, and the third and fourth planets trapping into a 5:4 resonance. In the 4-planet case the outer two planets locked into a 3:2 resonance.

In order for the three planets to finish at their appropriate initial semimajor axes, and to avoid numerical artifacts arising from abruptly turning off the migration, we chose to smoothly remove it by varying the exponential semimajor axis damping timescale $\tau_a$ as
\begin{equation}
\tau_a(t) = \frac{\tau_{a0}}{a_3(t) - a_{3\rm ALMA}}
\end{equation}
where $a_{3\rm ALMA} = 64.2$ AU and $\tau_{a0} = 90$ Myr (slow enough for the planets to capture into resonance) for the 5-planet case, and $a_{3\rm ALMA} = 69.0$ AU and $\tau_{a0} = 105$ Myr for the 4-planet case.  Thus, as the inner planet approaches its appropriate semimajor axis, the migration timescale smoothly diverges, effectively shutting itself off.  The simulation time was then reset to zero, and from that point treated like the non-resonant cases. 

For these setup simulations, we set the planetary masses and eccentricity damping timescales to constant values of $10 M_\oplus$ and 1 Myr, respectively (i.e., the initial values in the actual integrations, see Sec.\:\ref{setup}).  

\par The above procedure resulted in initial semimajor axes that matched the ALMA observations to within $\lesssim 1 AU$, which is smaller than the size of the synthesized beam in the ALMA data \citep{Brogan15}. The resulting initial eccentricities for the resonantly interacting planets were $\sim 10^{-3}$, while those for the dynamically detached inner two planets were $\sim 10^{-5}$ . A summary of the initial conditions for each case can be found in Table \ref{tableconditions}.

\section{Results} \label{results}

\par We begin by comparing our results following 5 Gyr of evolution, both to one another and to the observed exoplanet sample. Table \ref{planetloss} summarizes the average number of ejected planets, planet-planet collisions (C), number of planets reaching inward of 0.2 AU (S) and number of remaining planets (R) for each of the four cases considered.

\begin{figure}
\includegraphics[width=\columnwidth]{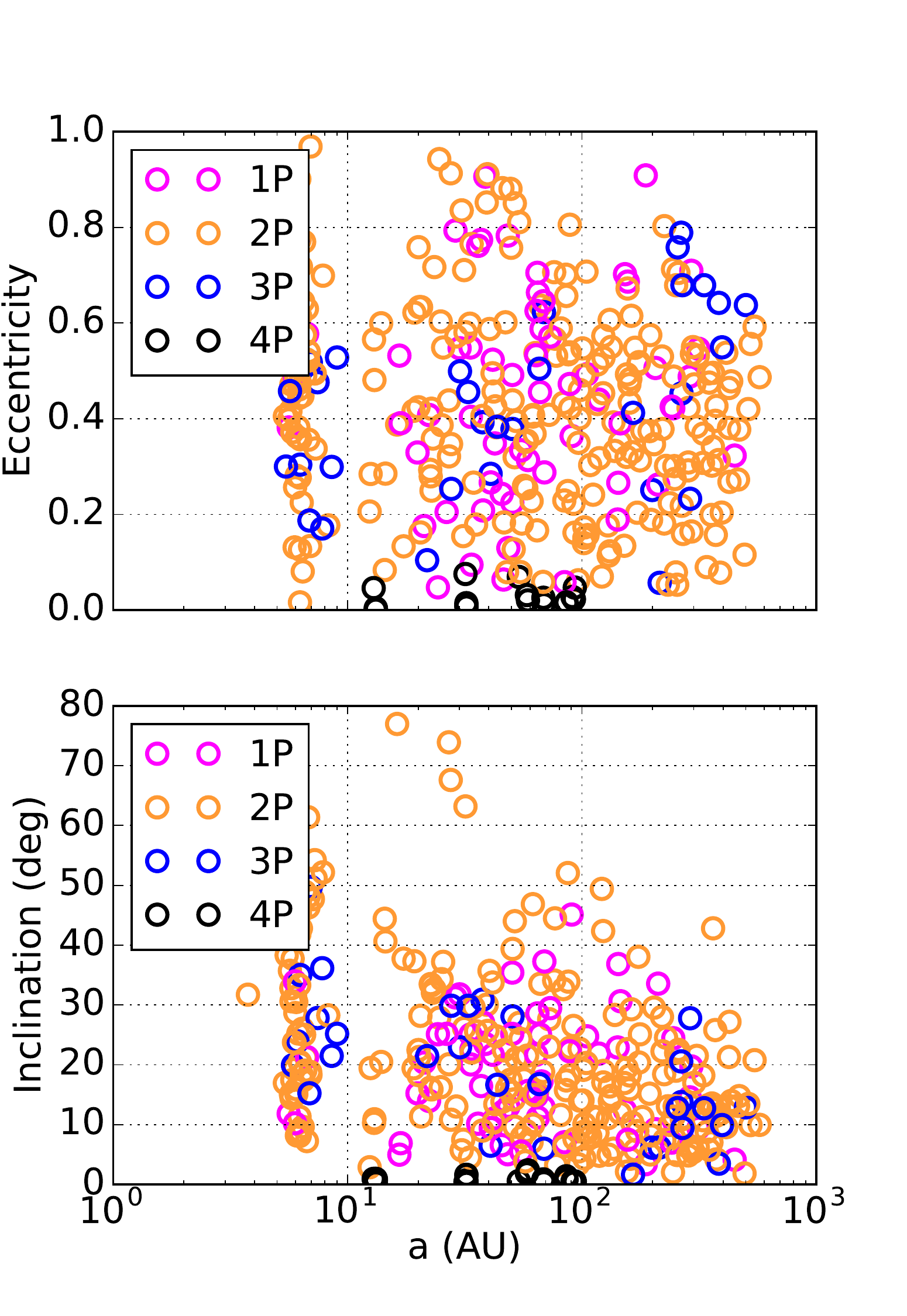}
\caption{Final semimajor axis vs. eccentricity (top) and semimajor axis vs. inclination relative to the initial plane (bottom) for the 5 planet cases. The different colors are used to indicate how many planets were left over in the system.}
\label{scatter}
\end{figure}

Figure \ref{scatter} shows a scatter plot of the final eccentricities and inclinations (relative to the initial plane) vs. the semimajor axes, color-coded by the number of remaining planets. The inner limit at $\approx 6$ AU is set by conservation of energy, where the innermost planet scatters outward or ejects the remaining planets, absorbing their (negative) orbital energy.

We find that in a subset of cases (black circles in Fig.\:\ref{scatter}), a planet is ejected while leaving the remaining planets in a stable, long-lived configuration. However, in the majority of cases, the eccentricities are raised to a level where the planets continue to vigorously interact and the system relaxes to a steady distribution \citep{Chatterjee08, Juric08} with a wide range of eccentricities and inclinations.

In the four-planet cases, most ejections and collisions occurred within 40 Myr ($\sim 10^4$ outer planet orbits), with a tail extending to Gyr timescales. The distribution was broader in the five-planet cases, where ejections more often left the remaining planets on orbits that would continue to interact strongly.

\begin{table}
\centering
\begin{tabular}{|l|c|c|c|c|}
\hline \hline
Case                  & E    & C    & S    & R    \\ \hline \hline
5 Planet Resonant     & 2.39 & 0.19 & 0.75 & 1.67 \\ \hline
5 Planet Non-Resonant & 2.41 & 0.07 & 0.68 & 1.84 \\ \hline
4 Planet Resonant     & 1.68 & 0.05 & 0.24 & 2.03 \\ \hline
4 Planet Non-Resonant & 1.45 & 0.05 & 0.27 & 2.23 \\ \hline
\end{tabular}
\caption{The final average number of planets lost to ejections (E), planet-planet collisions (C), close encounters with the star at 0.2 AU (S), and the final average number of planets remaining (R).}
\label{planetloss}
\end{table}

\subsection{Free-floating Planets} \label{freefloating}
\par Statistics of observed free-floating planets from \cite{sumi2011} found there to be $1.8^{+1.7}_{-0.8}$  Jupiter mass objects per star, of which they estimate 75\% of them are unbound. This corresponds to $1.3^{+1.3}_{-0.6}$ free-floating planets per main-sequence star. Similar estimates were obtained from a recent synthesis of radial velocity, microlensing and direct imaging survey results \citep{Clanton16}.

While our numerical results show planets are ejected with high efficiency (1.5-2.4 planets per star system, Table \ref{planetloss})\footnote{We note that one would expect to obtain similar results for unstable planetary systems at smaller semimajor axes, up to the point where relative velocities between planets become comparable to the escape velocity from the largest body (scales of $\sim 1 AU$ for Jupiter mass planets around a solar mass star). Closer in, collisions become the dominant outcome (rather than ejections), which would drive the expected rate of free-floating planets down further.}, one must additionally fold in the fraction of stars that host giant planets in the first place. \cite{mayor2011harps} estimate that 13.9 $\pm 1.7 \%$ of FGK stars host giant planets (larger than 50 Earth masses) with orbital periods under 10 years \citep[see also][]{Cumming08}. Therefore, while planet-planet scattering of giant planets efficiently ejects them, it would seem that there are too few gas giants formed in nature to match the rate of free-floating planets inferred from microlensing surveys \citep[see also][]{Veras12}. The upcoming WFIRST-AFTA mission should detect many more unbound planets and help clarify this situation \cite[e.g.][]{Spergel13}.

\subsection{Sensitivity to Initial Conditions} \label{sensitivity}
\begin{figure}
\includegraphics[width=\columnwidth]{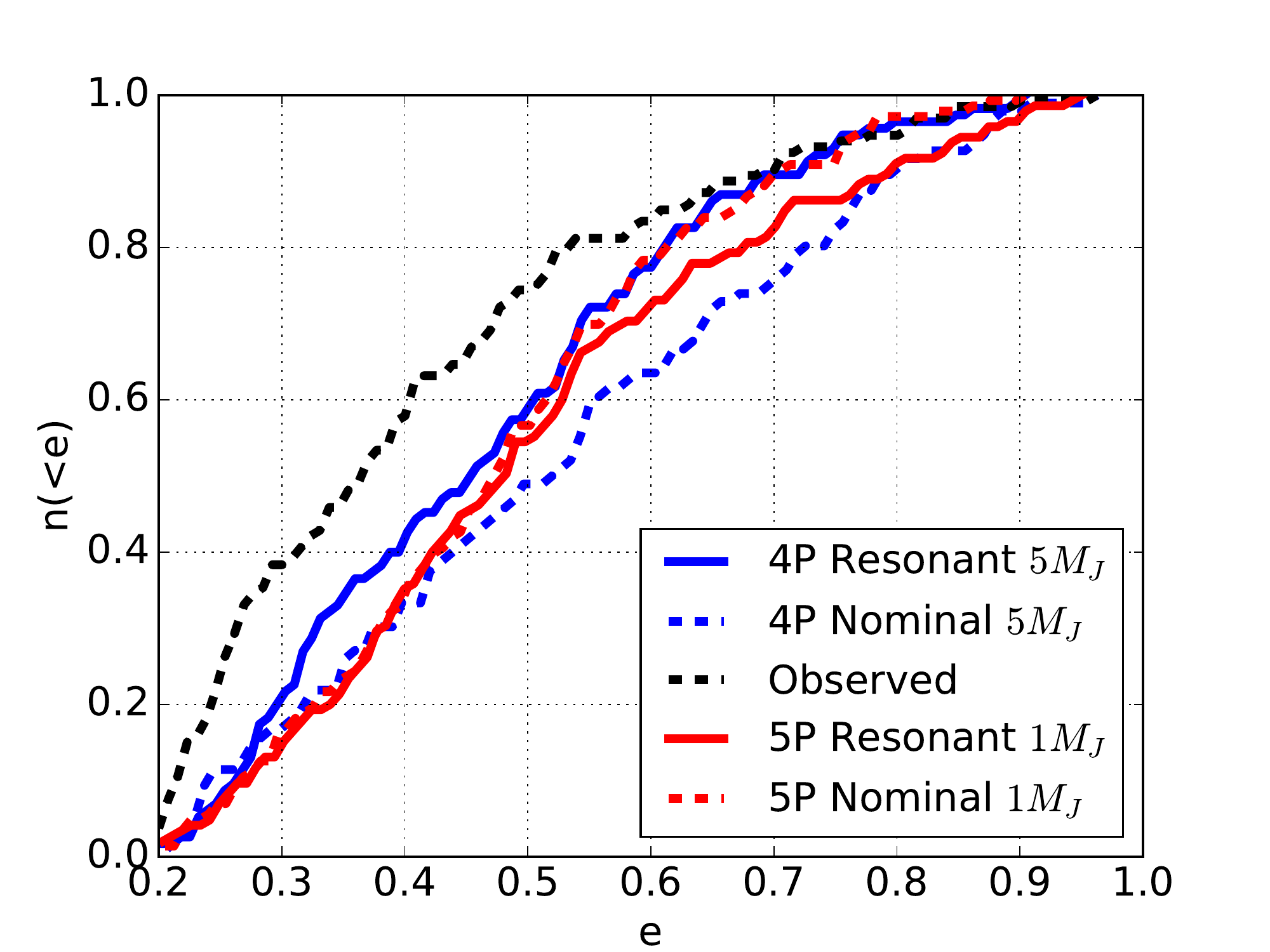}
\caption{Cumulative distribution function of eccentricity for the simulated results of each of the cases listed in Table \ref{tableconditions}, as well as the observed sample. Each cumulative distribution only contains planets with $e \geq 0.2$. See the main text for the observed sample used to generate the black dashed line}
\label{cdfe4p5p}
\end{figure}

To compare the outcomes from our four suites of simulations, we plot the cumulative distributions of the remaining planets' final eccentricities after 5 Gyr of evolution in Figure \ref{cdfe4p5p}. As mentioned above, not all systems (16 in the resonant case and 27 in the non-resonant case) in the four-planet cases became unstable.  In order to compare to both the five-planet cases (always unstable) and to the observed sample (see Sec.\:\ref{comptoobserved}), we only consider planets with eccentricities $> 0.2$.

\par Figure \ref{cdfe4p5p} shows that the overall distribution of outcomes is not particularly sensitive to initial conditions.  We performed Kolmogorov-Smirnov (KS) tests between all the possible pairs of cases in Table \ref{tableconditions}. These revealed p-values as low as 0.11 (between the 4-planet nominal and resonant cases) and as high as 0.69 (between the 5-planet nominal and resonant cases). We therefore can not rule out that the null hypothesis that they were drawn from the same underlying distribution. This relative insensitivity to initial conditions is consistent with the results of previous planet-planet scattering experiments \citep[e.g.,][]{Chatterjee08, Juric08}. We also note that while planets' final distances from the star will depend on their detailed migration histories, this insensitivity to initial conditions suggests that planet-planet scattering will erase migration signatures in the eccentricities (and inclinations) of unstable planetary systems, relaxing to a dynamically imposed equilibrium distribution.

\subsection{Comparison to the Observed Exoplanet Sample} \label{comptoobserved}

Our results can also be compared to the observed exoplanet sample. While the HL Tau system has a scale ($\sim 100$ AU) probed by direct imaging, it is difficult to make a direct comparison due to the comparatively small number of observed planets, and the limited constraints on their orbital eccentricities and inclinations (due to the typically short observed orbital arcs). 

However, because point-source Newtonian gravity is scale-free, we can consider the HL Tau system as a scaled up prototype of initial conditions at $\sim 1$ AU and compare our results to the observed sample of radial velocity (RV) giant planets (i.e., we would have obtained similar results for dimensionless quantities like the orbital eccentricities and inclinations if we had scaled all our semimajor axes down by a constant factor). In this picture, our 5 Gyr integrations represent $\approx 10^8$ inner planet orbits, which is short compared to RV systems' ages, but long enough for the resulting orbital distributions to approximately converge \cite{Chatterjee08, Juric08}. 

We note, however, that our results are not strictly scale-free, since we consider collisions and our bodies are therefore not point particles. The importance of finite planetary and stellar radii for collisions varies as we scale down the system from tens of AU to $\sim 1$ AU. This is  negligible for the planetary radii, since interplanetary collisions are rare (we observed 63 across 1800 planets in 400 simulations). In the case of the star, we inflated its radius to 0.2 AU.  When taken as a fraction of the system size $\sim 50 AU$, this corresponds to the size of a sun-like star for a system with characteristic semimajor axes of $\sim 1 AU$, which makes our simulations comparable to previous studies \citep{Chatterjee08, Juric08}. Technically, our addition of mass growth and eccentricity damping also introduce new scales into the problem, but as argued in Sec.\:\ref{setup} these have minimal effect on the dynamics of this particular problem.


In order to compare to the observed sample, we drew data from \url{exoplanets.org} on planets discovered through RV with 1 AU$<$ a $<$ 5 AU, and with $e > 0.2$.  This should predominantly select systems that underwent some form of dynamical excitation, and are thus candidate outcomes for (scaled down) initial conditions like those in HL Tau.

From Fig.\:\ref{cdfe4p5p}, we see that the observed sample has significantly lower eccentricities than our simulations provide.  KS tests between our various simulated cases and the observed distribution yield p-values $\lesssim 10^{-3}$.  This suggests that equal-mass planets with HL Tau-like initial conditions can not reproduce the observed eccentricity distribution. We therefore add next a distribution of planetary masses.

\subsection{Adding a Mass Distribution} \label{massdistribution}

Typically, scatterings between equal-mass planets will affect both orbits comparably. By contrast, if one planet is much more massive than the other, the diminutive body can achieve large eccentricities while only mildly perturbing the massive one. Adding a distribution of planet masses can therefore shift the cumulative eccentricity distribution closer to what is observed \citep{ford2008, Chatterjee08, Juric08}.  

To explore this, we duplicated the initial conditions from the nominal five-planet case (Table \ref{tableconditions}), and ran two additional suites of simulations, one generated by drawing $\tau_{M0}$ (Eq.\:\ref{taum}) from a uniform distribution on the interval [0.70, 1.30] Myr, the other on the interval $[0.65$  Myr$, 1.30 $ Myr$]$ respectively. These mapped (non-uniformly) onto a mass distribution between [0.3, 2.3] $M_J$ and [0.3, 3.2] $M_J$, respectively. 

\begin{figure}
\includegraphics[width=\columnwidth]{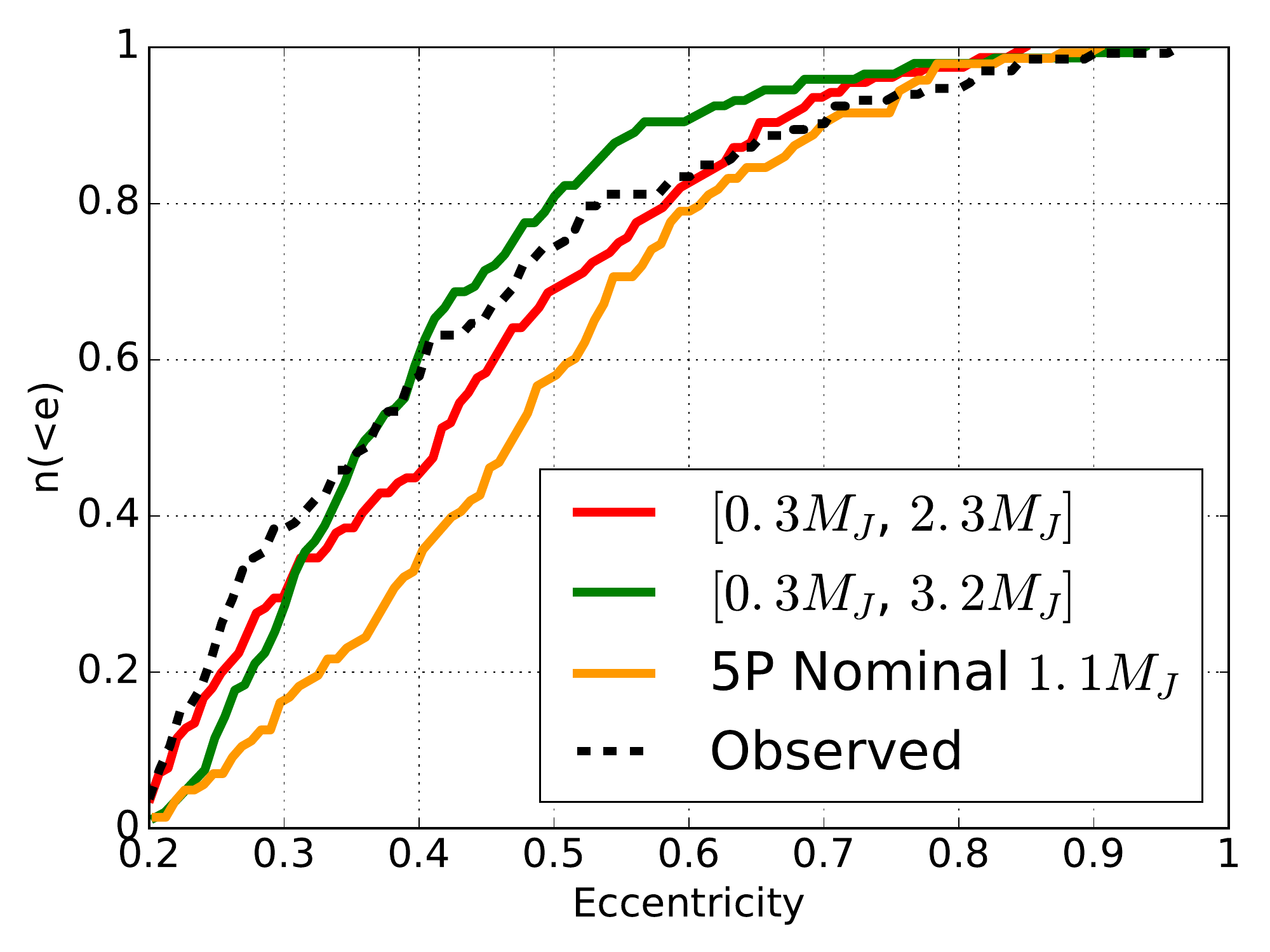}
\caption{Cumulative distribution function of eccentricity for the systems with varying mass ratios, the observed sample, as well as the five planet non-resonant case from Fig.\:\ref{cdfe4p5p} for reference. The samples contain only planets with $e \geq 0.2$}
\label{cdf5peMV}
\end{figure}

The resulting cumulative eccentricity distributions can be seen in Fig.\:\ref{cdf5peMV}, along with the equal-mass nominal five-planet case. We observe that by increasing the range in masses, one generates a higher-proportion of low eccentricity planets (i.e., the massive ones that are less perturbed during encounters).  By tuning the mass distribution it is thus possible to obtain a closer match to the observed population.

The best fit to the observed sample is the case with masses between [0.3,3.2] $M_J$. A K-S test reveals a 63\% probability of equal or worse disagreement between the two distributions under the hypothesis that they are in fact drawn from the same distribution.  Thus, scaled down HL Tau initial conditions with a plausible range of masses can reproduce the observed eccentricity distribution.



\subsection{Inclinations} \label{inclinations}
While mutual planetary inclinations have only been measured for a handful of systems
\citet{mcarthur10,dawson14,MF16}, GAIA is expected to constrain the mutual inclinations between large numbers of giant planets at $\sim 1$ AU distances \citep{casertano08,perryman}. We therefore consider the inclination distribution generated by planet-planet scattering. Previous authors have typically presented predictions for the distribution of absolute inclinations to their respective system's invariable plane \citep[see, e.g., Fig 12 in][]{Juric08}; however, this does not accurately reflect the distribution of mutual inclinations between pairs of planets.

\begin{figure}
\includegraphics[width=\columnwidth]{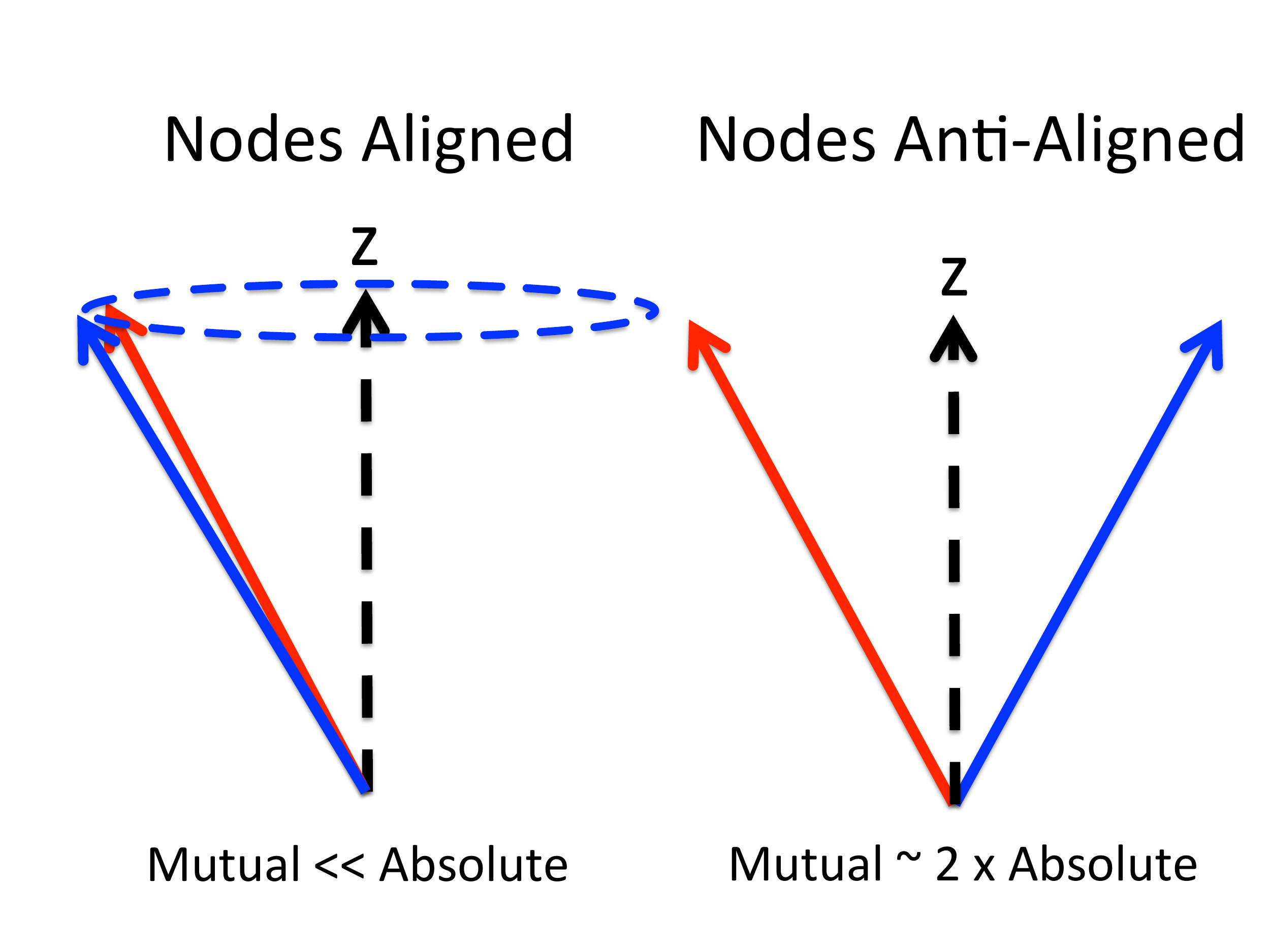}
\caption{Orbit normals for a pair of planets with approximately equal absolute inclinations, with aligned and anti-aligned nodal configurations. Any value is possible for the mutual inclination between 0 and twice the absolute inclination, depending on where the pair of orbit normals lie on the cone traced out by the dashed blue circle.}
\label{incdiagram}
\end{figure}

Consider two orbits with the same absolute inclination (i.e., the angle between the orbit normal and $z$ axis, defined by the invariable plane). As shown in Fig.\:\ref{incdiagram}, these can have orbit normals pointing in many different directions that lie along a cone. Thus, two orbits with the same inclination might have zero mutual inclination if their orbit normals coincide, but their longitudes of node (i.e., the azimuthal angle along the dashed blue circle in Fig.\:\ref{incdiagram}) could also be $180^\circ$ apart, corresponding to a mutual inclination that is twice as large as their absolute inclinations (one could also draw any configuration in between these two extremes). But if the two planets make up a closed system, then the total angular momentum must always lie along the invariable plane's axis $z$, and only the anti-aligned case is possible. This simple picture is complicated by planets having different absolute inclinations, the presence of additional planets, and the fact that ejected planets can permanently remove angular momentum from the system; nevertheless, one expects the distribution of mutual inclinations between planet pairs to be wider than the distribution of absolute inclinations.

\begin{figure}
\includegraphics[width=\columnwidth]{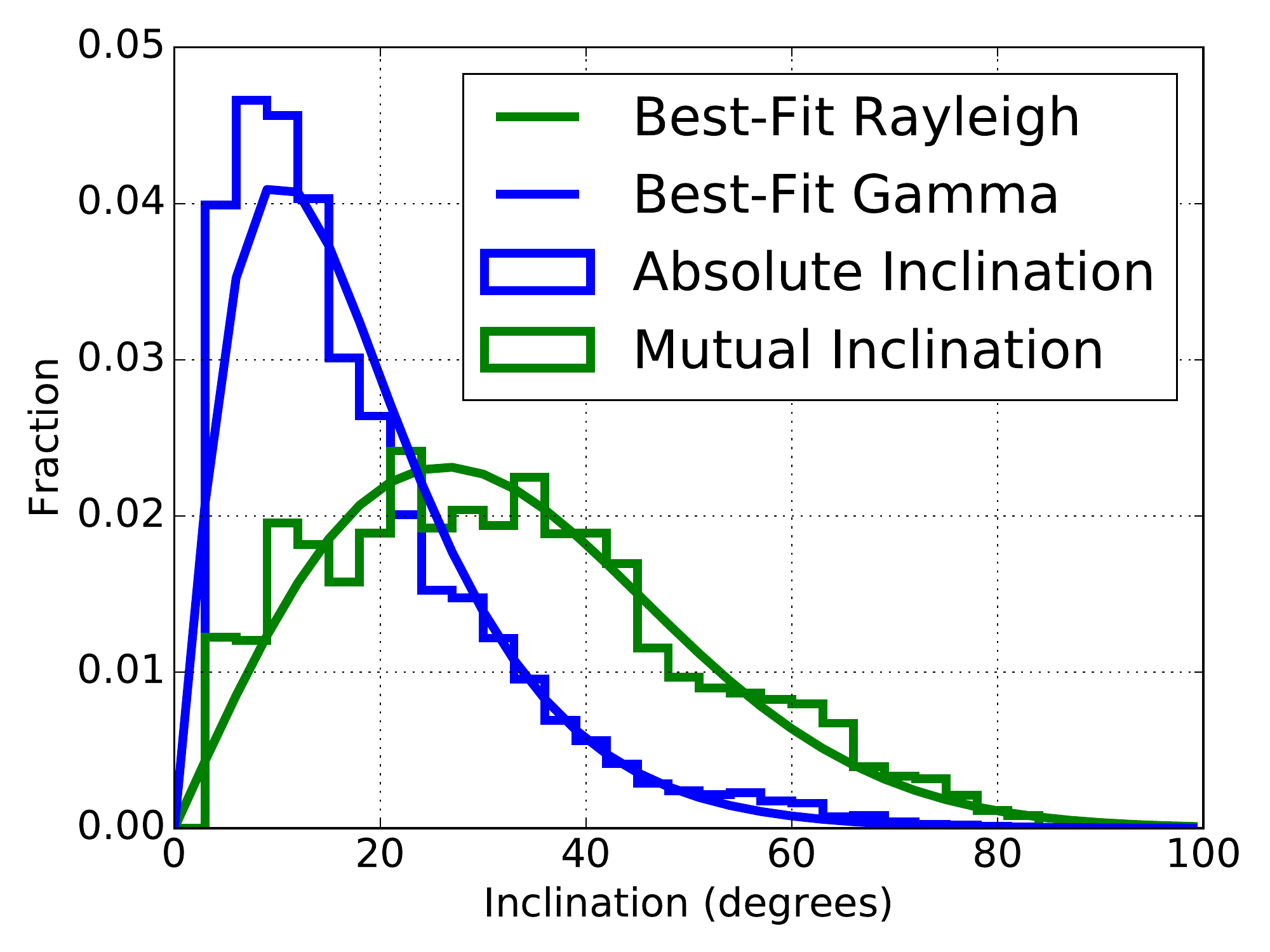}
\caption{Normalized distribution of absolute and mutual inclinations with all cases in Table \ref{tableconditions} combined in a single dataset.}
\label{hist_inclinations}
\end{figure}

Figure \ref{hist_inclinations} compares the distributions of absolute and mutual inclinations in our numerical experiments, where we have manually removed the smallest-inclination bin in order to remove stable systems from consideration.
Given the insensitivity to initial conditions (Sec.\:\ref{sensitivity}), we combined the four suites of simulations listed in Table \ref{tableconditions} to generate the resulting distributions.  We see that indeed, the mutual inclination distribution is significantly wider.  Because our results are largely scale-free (Sec.\:\ref{comptoobserved}), they provide predictions for HL Tau-like initial conditions at the $\sim 1$AU scales that will be probed by GAIA (e.g., \citealt{casertano08}). 

We find through maximum likelihood estimation that the mutual inclination distribution is well fit by a Rayleigh distribution with a scale parameter of $26.2^\circ$ ($26.1^\circ-26.4^\circ$ at 95\% confidence). Rayleigh distributions could not match the absolute inclinations, but they are reasonably fit by a Gamma distribution with shape parameter of $2.3$ (2.26-2.34 at 95\% confidence) and scale parameter of $7.9^\circ$ ($7.8^\circ-8.1^\circ$ at 95\% confidence).

As shown by \citet{timpe13}, similar to the eccentricity distribution (Sec.\:\ref{massdistribution}), the distribution of mutual inclinations shrinks if we consider unequal-mass planets. Thus, our distribution can be regarded as maximal. 

\subsection{Production of Hot Jupiters} \label{hotjupiters}

\subsubsection{Numerical Setup}

\par In order to explore the production of hot Jupiters in our simulations, we reloaded simulations in the two resonant cases in Table \ref{tableconditions}, at the first moment that a planet reached 0.2 AU from the central star.

We then added precession effects from general relativity and the tides raised by the star on the planets. We added general relativity using the {\tt gr} implementation in {\tt REBOUNDx}, which adds the first post-Newtonian approximation (1PN), i.e. terms quadratic in the ratio of the velocities to the speed of light, but ignores terms that are smaller by a factor of the planet-star mass ratio \citep{Anderson75}.  We used the {\tt tides\_precession} implementation in {\tt REBOUNDx} for the precession induced by the interaction between the central body and the tidal quadrupoles raised on the planets by the star, which is based on \cite{Hut81}. We adopted an apsidal motion constant $k_1$ (half the tidal Love number) of 0.15 for all planets.

Mostly because in this case the IAS15 integrator must resolve the very short periastron passages, each of the two 100-simulation cases we explored required $\sim 10^5$ CPU hours of integration time using the same hardware. Because of the computational cost, and the insensitivity to whether or not planets are started in resonance (Sec.\:\ref{sensitivity}), we only ran the four and five-planet resonant cases (Table \ref{tableconditions}) for this analysis.

\subsubsection{Rates}
\par The planets that approach very close to their host stars can be strongly affected by tides and circularize their orbits giving rise to a population of the close-in planets ($a\lesssim0.1$ AU), the so-called hot Jupiters (e.g., \citealt{rasio1996,nagasawa2008}). 

We have ignored the effect of this tidal dissipation in our calculations mainly because the formation of short-period planets demand extremely short integration time-steps. Instead, we have included only the conservative potential describing the interaction of the star with the tidal bulge it raises on the planets \citep{Hut81}, but have kept track of the minimum distance $r_{\rm min}$ between each planet and the host star. This quantity is a good proxy to determine whether or not a planet will tidally capture because the orbital circularization timescale depends very steeply on $r_{\rm min}$, so migration should occur below a certain threshold. Because planet migration should proceed at nearly constant orbital angular momentum ($a[1-e^2]\simeq {\rm constant}$), a planet reaching an $r_{\rm min}=a(1-e_{\rm max})$ that is small enough for tides to damp its orbital eccentricity within $5$ Gyr will reach a final semimajor axis $a_{\rm f}\simeq a(1-e_{\rm max}^2) = r_{\rm min}(1+e_{\rm max})\simeq 2r_{\rm min}$.

\begin{figure}
\includegraphics[width=\columnwidth]{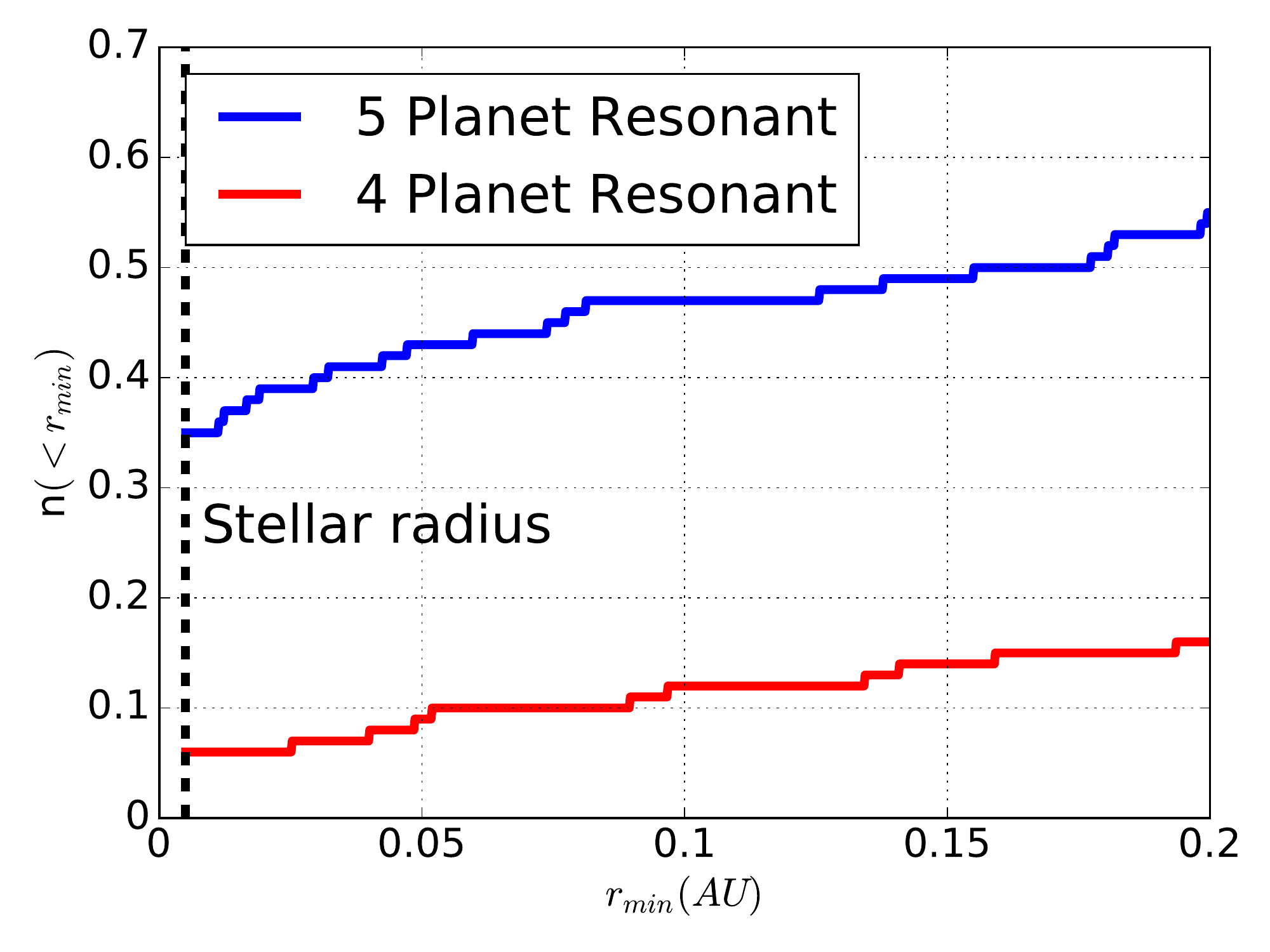}
\caption{Cumulative distribution function of the minimum distance from the host star that was reached by the closest remaining planet, for both the four and five-planet resonant cases (Table \ref{tableconditions}). The fraction plotted at the stellar radius of $\approx 0.005 AU$ collided with the central star.}
\label{fig:rmin}
\end{figure}

In Fig.\:\ref{fig:rmin} we plot the cumulative distribution of $r_{\rm min}$ for the four and five-planet resonant cases (Table \ref{tableconditions}). Depending on the efficiency of tides, we can determine the fraction of planets that can potentially become a hot Jupiter. For instance, the dynamical tides model by \citet{IP07,IP11} predicts that the orbits of Jupiter-like planets can be efficiently circularized when $r_{\rm min}\lesssim0.03\mbox{ AU}\simeq 6R_\odot$ and, therefore, form hot Jupiters with semi-major axes $a_{\rm f}\lesssim0.06$ AU.

We see from Table \ref{planetloss} that while nominal vs resonant initial conditions make little difference in the number of planets coming close to their host star (column S), there is a significant difference between the four and five-planet cases. This is also evident in Fig.\:\ref{fig:rmin}. Significant dependencies on the number of planets in the system have also been found in previous studies \citep[e.g.,][]{Juric08}. We note that while the final eccentricity distribution is largely scale free (Sec.\:\ref{comptoobserved}), asking what fraction of planets reach pericenter distances at which tides can operate introduces a length scale into the problem, rendering initial conditions important. 

We also note that if the eccentricity evolution were purely secular, and thus slowly varying, the precession induced by the interaction between the tidal bulge induced on the planets and the star would cause the pericenter to stall at a minimum distance of $\sim 3$ stellar radii, or 0.015 AU \citep{Wu11}. However, due to the strong interactions between planets, we find that a large fraction impact the star, despite this strong tidal precession (35\% and 6\% in the five and four-planet cases, respectively). However, one would expect that if we had included tidal dissipation, most of these planets would have captured to form hot Jupiters before colliding with the central star.

In total, we find that in 40\% and 7\% of the five and four-planet systems, respectively, a planet passes within our adopted threshold of 0.03 AU of the central star \citep{IP07,IP11}, and is thus a candidate for capture as a hot Jupiter. As discussed above, their final semimajor axes will be $\lesssim 0.06$ AU, but because we do not consider tidal dissipation, we cannot predict the distribution of semimajor axes within this range. The above percentages are also likely  overestimates, since \citet{guillochon11} predict that a Jupiter-like planet orbiting a Sun-like star will be disrupted if it passes within $\approx 0.01$ AU of the central body, and dissipation in the host star can pull close-in planets into the star \citep{BN12}. Our results highlight the importance of initial conditions, but are broadly consistent with \citet{nagasawa2008,nagasawa2011}, who predict that scattering of three-equal-mass planets (with the innermost body initially at 5 AU) form hot Jupiters in $\sim30\%$ of the systems, and \citet{BN12}, who generated hot Jupiters in $\sim10$ and $\sim20\%$ of their three and four-planet systems (innermost body initially between 1-5 AU), respectively.

Putting together our hot Jupiter formation rates with the fraction of FGK stars hosting giant planets \citep[$\sim 10\%$][]{mayor2011harps,Cumming08} yields an expected rate of $\sim 0.7-4\%$ hot Jupiters per FGK star. This brackets the rate found by radial velocity surveys of $1.2 \pm 0.38 \%$ \citep{wright2012}, with the caveat that our estimates would be lowered by tidal disruptions \citep{guillochon11} and hot Jupiters migrating into the host star through tidal dissipation in the central body \citep{BN12}. Since the above studies have considered these effects in detail, we do not pursue them further.




\subsubsection{Stellar Obliquities}

We finally briefly discuss the inclination distribution of planets as they approach the host star at high eccentricities. In particular, this inclination distribution at closest approach can be significantly different than the inclination distribution over all times. 

To track this, we recorded the inclinations (with respect to the initial invariable plane) of the planets at the time of closest-approach. We expect that, given the steep dependence of the tidal dissipation rate on $r_{\rm min}$, the inclination at closest approach should be a good proxy for the final inclination of a hot Jupiter. If we additionally assume that the initial stellar obliquities are small (the equator is nearly aligned with the system's invariable plane), then these planetary inclinations closely correspond to the star's final obliquity to the hot Jupiter's orbital plane, or the spin-orbit misalignment.

\begin{figure}
\includegraphics[width=\columnwidth]{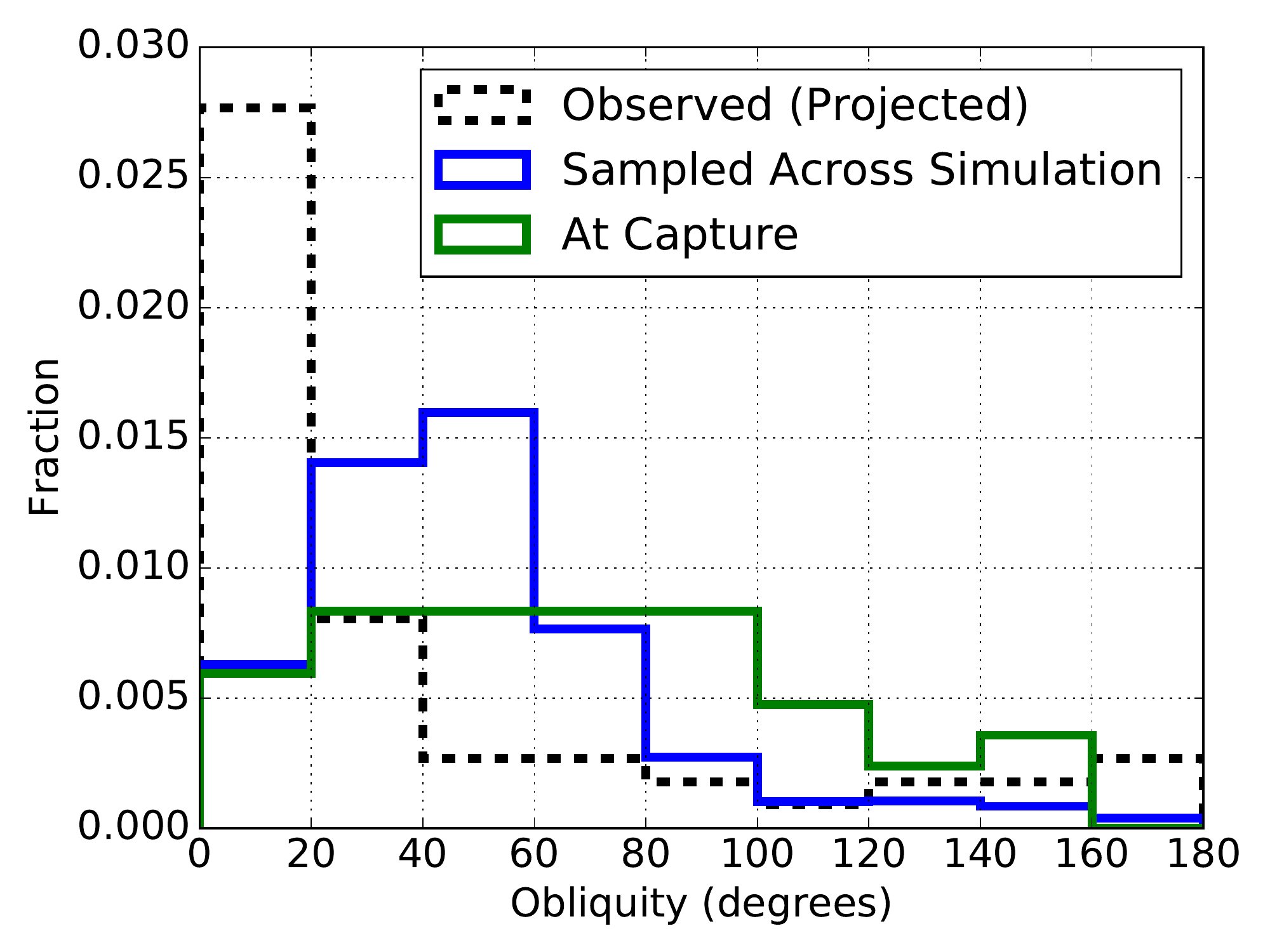}
\caption{Distribution of stellar obliquities (angle between the stellar spin and planet's orbital plane) for planets reaching within 0.1 AU of the host star, in the resonant cases listed in Table \ref{tableconditions}. We assume that the stellar spin remains aligned with the system's invariable plane. Secular effects cause the obliquities to have a significantly flatter distribution at high eccentricities where the planets are captured to make hot Jupiters (green distribution) than they do over all phases of the secular evolution (blue distribution). The expected hot Jupiter obliquity distribution (green) does not match the black dashed distribution of observed obliquities (projected onto the plane of the sky).}
\label{fig_obliquities}
\end{figure}

In Figure \:\ref{fig_obliquities} we show in green the distribution of stellar obliquities for planets that come close to their host star. In order to increase our statistics, we combine the results from our four and five-planet resonant cases, and consider any planet that reached within $r_{\rm min}<0.1$ AU from the central body. This distribution is significantly flatter than the blue distribution of planetary inclinations sampled at many times over each simulation. This indicates that the highest inclinations are reached during the episodes of extremely large eccentricities, possibly linked to strong scattering events and/or secular chaotic effects. 

We find that $\approx 26\%$ of planets reach inside $0.1 AU$ on retrograde orbits (obliquities $>90^\circ$). This is consistent with the $\sim30\%$ of retrograde hot Jupiters found by \citet{nagasawa2008,nagasawa2011} in their simulations.

If hot Jupiters are largely formed through planet-planet scattering, this predicts that hot Jupiters should have outer planetary companions. While earlier studies suggested hot Jupiters were found in single-planet systems \citep{Wright09, Steffen12}, recent work suggests they have similar companion fractions to giant planets farther out \citep{Schlaufman16}.

However, the expected flat distribution of hot Jupiter obliquities is inconsistent with observations (black dashed distribution in Fig.\:\ref{fig_obliquities}, taken from \url{exoplanets.org}), which show a strong preference for lower values, and a rate of retrograde systems of $\sim15\%$ (e.g., \citealt{Winn15}).
This suggests that if planet-planet scattering is responsible for a large fraction of the observed hot Jupiters, then tidal dissipation has played an important role in aligning the axes of the stellar spin and planet's orbit (see, e.g. \citealt{lai12,VR14,Dawson14b}).

\section{Conclusion}
The HL Tau system potentially provides the first set of initial conditions for planets as they emerge from their birth disks.
We have followed the evolution of such initial conditions for 5 Gyr using N-body integrations to explore the possible outcomes.  
Because the results are largely scale free down to $\sim 1 AU$ scales, where interplanetary collisions start to become important \citep[e.g.][]{Petrovich14}, they not only provide predictions for wide separation systems like HL Tau, but can also be directly compared to the observed sample of giant planets from radial velocity surveys.

We find that HL Tau initial conditions naturally produce several populations in the observed exoplanet sample:
\begin{itemize}[leftmargin=*]
	\item Eccentric cold Jupiters: We can match the observed eccentricity distribution of dynamically excited radial velocity planets with $e>0.2$, in agreement with previous planet-planet scattering studies \citep[e.g.,][]{Chatterjee08, Juric08}; we note that this result is sensitive to the unconstrained distribution of planetary masses. 
    \item Hot Jupiters: We obtain upper limits of $\sim 7-40\%$ for the production efficiency of hot Jupiters. When combined with the observed fraction of systems that host giant planets ($\approx 14\%$, \citealt{mayor2011harps}), this brackets the $\approx1\%$ rate of hot Jupiters observed around FGK stars \citep{wright2012}. Furthermore, we find that secular interactions lead to a significantly broader distribution of hot Jupiter obliquities than one would naively expect from the orbital inclination distribution from scattering experiments that ignore tidal dissipation.
    \item Free Floating Planets: We find a high efficiency for ejections of $\approx 2$ planets per HL Tau-like system. However, given the small fraction of systems that host giant planets (at least inside $\sim 5$ AU, \citealt{mayor2011harps}), it does not seem possible to match the high rate of free-floating planets inferred from microlensing surveys ($\sim 1$ per main sequence star, \citealt{sumi2011}).
\end{itemize} 

We also present an expected distribution of {\it mutual} inclinations between planets, which will be probed by the GAIA mission. We find that due to angular momentum conservation, it is significantly broader than the distribution of absolute inclinations often reported by planet-scattering studies.

This study is consistent with previous work on planet-planet scattering \citep[e.g.,][]{Juric08, Chatterjee08} showing that dynamically unstable systems lose memory of their initial conditions and relax to equilibrium orbital distributions. This makes it possible for initial conditions from a single system (HL Tau) to lead to features observed in the exoplanet sample as a whole. The striking observation is that the HL Tau gaps plausibly correspond to planets that are long-lived relative to the age of the system \citep{Tamayo15}, but are destined for dynamical instabilities on longer timescales that can produce many of the evolved systems we see today.

\section*{Acknowledgements}

We would like to thank Yanqin Wu and Phil Armitage for insightful discussions. We are also grateful to the anonymous referee who greatly helped improve and sharpen this manuscript. D.T. is grateful for support from the Jeffrey L. Bishop Fellowship. H.R. was supported by NSERC Discovery Grant RGPIN-2014-04553. C.P. is grateful for support from the Gruber Foundation Fellowship. This research was made possible by the kind and tireless support of Claire Yu and John Dubinski, and by the Sunnyvale cluster at the Canadian Institute for Theoretical Astrophysics. This work was greatly aided by the open-source projects \texttt{Jupyter} \citep{jupyter} \texttt{iPython} \citep{ipython}, \texttt{SciPy} \citep{scipy} and \texttt{matplotlib} \citep{matplotlib, matplotlib2}.



\bibliographystyle{mnras}




\bsp	
\label{lastpage}
\end{document}